# Active Coding Piezoelectric Metasurfaces

Zhaoxi Li,[1] Chunlong Fei,[1*] Shenghui Yang,[1] Chenxue Hou,[1] Jianxin Zhao,[1] Yi Li,[1] Chenxi Zheng,[1] Heping Wu,[3] Yi Quan,[1,3] Tianlong Zhao,[1] Dongdong Chen,[1] Di Li,[1] Gang Niu,[3] Wei Ren,[3] Meng Xiao,[2*] and Yintang Yang[1*]

[1]School of Microelectronics, Xidian University, Xi'an 710071, China

[2]Key Laboratory of Artificial Micro- and Nano-structures of Ministry of Education and School of Physics and Technology, Wuhan University, Wuhan 430072, China

[3]Electronic Materials Research Laboratory, Key Laboratory of the Ministry of Education & International Center for Dielectric Research, School of Electronic Science and Engineering, Xi'an Jiaotong University, Xi'an 710049, China

Abstract:

The manipulation of acoustic waves plays an important role in a wide range of applications. Currently, acoustic wave manipulation typically relies on either acoustic metasurfaces or phased array transducers. The elements of metasurfaces are designed and optimized for a target frequency, which thus limits their bandwidth. Phased array transducers, suffering from high-cost and complex control circuits, are usually limited by the array size and the filling ratio of the control units. In this work, we introduce active coding piezoelectric metasurfaces; demonstrate commonly implemented acoustic wave manipulation functionalities such as beam steering, beam focusing and vortex beam focusing, acoustic tweezers; and eventually realize ultrasound imaging. The information coded on the piezoelectric metasurfaces herein is frequency independent and originates from the polarization directions, pointing either up or down, of the piezoelectric materials. Such a piezoelectric metasurface is driven by a single electrode and acts as a controllable active sound source, which combines the advantages of acoustic metasurfaces and phased array transducers while keeping the devices structurally simple and compact. Our coding piezoelectric metasurfaces can lead to potential technological innovations in underwater acoustic wave modulation, acoustic tweezers, biomedical imaging, industrial non-destructive testing and neural regulation.

# 1. Introduction

Sound wave manipulation is of central importance in many applications, from audible to ultrasonic and from airborne to underwater. The last decade has witnessed the success of a family of thin artificial materials, dubbed metasurfaces, in arbitrarily controlling both the phase and amplitude of acoustic waves in a passive way[1]. Acoustic metasurfaces consist of two-dimensional arrays of subwavelength units such as Helmholtz resonators and coiling-up space structures[2,3,4,5]. With the development of fabrication methods, especially high-precision 3D printing techniques, metasurfaces have demonstrated fascinating capabilities in various airborne sound applications, including anomalous reflection[6], vortex beams[7], nondiffracting beams[8], self-bending beams[9], holograms[10], near-perfect absorption[11] and many others[12,13,14]. Since the subwavelength units (pixels) exhibit a frequency-dependent response, the optimization of metasurfaces is typically conducted in a frequency-specific manner, and the complexity of the design increases significantly as more pixels are involved and when the mutual coupling between pixels is considered. Another dominant strategy for sound wavefront manipulation is through active phased array transducers (PATs), wherein each pixel is controlled independently in terms of both phase and amplitude[15,16]. PATs can be configured in real time, exhibit enhanced dexterity at the cost of requiring a sophisticated supporting electrical system, and have also demonstrated various sound wave manipulation capabilities[17,18]. However, due to challenges hindering the development of reliable processes for fabrication, synchronization, electrical interconnection and system integration, the number of independent control units in a PAT is limited.

In this work, we introduce wideband active piezoelectric metasurfaces (APMs) and experimentally demonstrate several common ultrasound manipulation capabilities, including beam steering, beam focusing, vortex beams, microparticle trapping and manipulation and ultrasound imaging. The control information is encoded onto such an APM via the polarization directions, pointing either up or down, of the piezoelectric materials for different pixels, and this binary information is frequency independent. Different from traditional PATs, our APMs need only one controlling electrode, and the filling ratio of piezoelectric materials can approach almost one. In contrast to the passive approach in which metasurfaces are placed at certain distances from the sources, in this case, pixels with different responses simply consist of piezoelectric materials that generate acoustic waves; hence, the whole system is extremely compact. Our work incorporates characteristics of both dominant approaches, i.e., metasurfaces and PATs, and points towards a succinct sound wave manipulation scheme that can be achieved with techniques that are favourable for massive-scale industrial production.

## 2. Result

### 2.1 . A Coding Piezoelectric Metasurface With Opposite Polarizations

As sketched in **Figure 1**a, an unpoled piezoelectric material consists of randomly oriented dipoles. The poling process endows a piezoelectric material with an aligned polarization direction via the application of a large electric field. (We use the piezoelectric material PZT-4 in this work, and the details of the poling process are provided in the Supplementary Materials, Sec. 1) Depending on the direction of the applied electric field, a piezoelectric material can exhibit two opposite bulk polarizations, as illustrated in orange and cyan. In the lower left (right) panel of Figure 1a, a surface of residual positive (negative) charge emerges on the upper boundary, while a negative (positive) charge appears on the lower boundary; thus, these poled piezoelectric materials will react oppositely under the application of a small electric field. As shown in Figure 1b, when a poled piezoelectric material is subjected to a small AC field, it will generate a longitudinal acoustic wave as the piezoelectric crystal is either compressed or stretched. Thus, the generated acoustic wave can be either in phase or out of phase with the applied AC field depending on the direction of polarization. We emphasize that such binary phase information is independent of the frequency or amplitude of the signal, which greatly simplifies the design. The direction of polarization offers a means of binary phase modulation in the generated acoustic field, which can be further encoded into piezoelectric materials to form an APM. Simple as it is, we will show that this kind of APM is able to demonstrate several vital ultrasound manipulation capabilities. Figure 1c and 1d show two typical setups for arranging this binary phase information. In Figure 1c, domains of different polarizations form a chessboard pattern, and the AC signal is applied to the APM by a single electrode. With this setup, the direction of the generated sound wave can be smoothly varied by changing only the driving frequency. In Figure 1d, domains of different polarizations form concentric annuluses, and this setup is capable of generating focused beams and vortex beams.

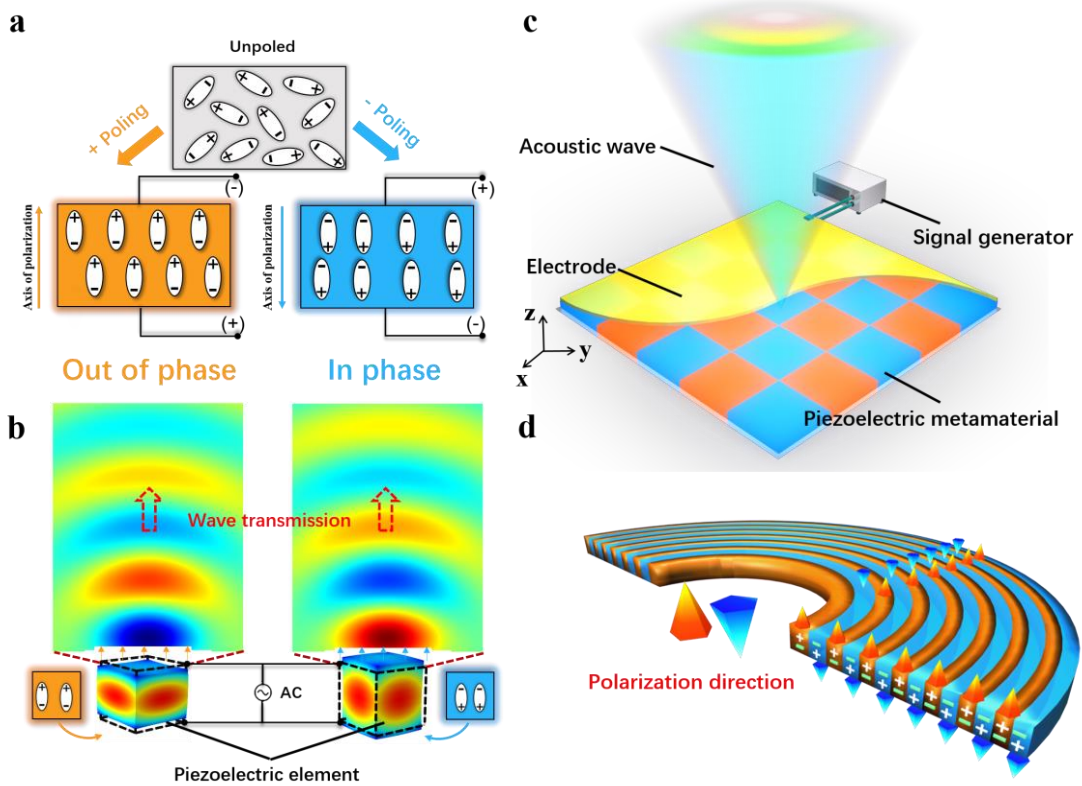

**Figure 1.** Binary information encoded in piezoelectric metasurfaces. **a,** The poling of piezoelectric crystals. **b**, Sound waves are generated as piezoelectric materials are compressed or stretched under small applied AC signals. Depending on the direction of the DC field applied during the poling process, poled piezoelectric materials can exhibit two opposite bulk polarizations, and the corresponding phases of the generated sound waves differ by π. Domains of oppositely poled piezoelectric materials can be arranged in a chessboard pattern (**c**) or in concentric annuluses (**d**).

## 2.2 Beam Steering Using APM

The responses of the subwavelength units in most metasurfaces are frequency dependent; hence, a typical metasurface is designed and optimized for a target frequency and exhibits good performance over only a narrow frequency range. A reduction in the phase variation from a continuous 2π range to values of only 0 or π can enhance the robustness of sound wave manipulation. We first consider a chessboard APM, as sketched in Figure. 1c, which has 5 pixels along each direction, where the size of each pixel is 1.5×1.5 mm$^2$. After poling, the APM is assembled as an ultrasound source through a standard process. (Details can be found in the Supplementary Materials, Sec. 2.) The APM is driven by a single electrode to generate ultrasound in water. Similar functionalities have been realized with microwave metasurfaces[19] and acoustic metasurfaces in air[20]; here, the binary phase information is independent of frequency, and hence, our APM can work over a much broader frequency range with no additional accessories.

Figure 2a and 2b show the amplitude and phase of the pressure field 2.5 mm above the APM at a working frequency of 1.5 MHz respectively. In the simulation, each pixel of the APM is treated as a plane wave source with a uniform amplitude and a phase of either 0 or π. (Details of the simulations are provided in the Methods section and in the Supplementary Materials, Sec. 3.) The pressure field patterns in the experiments and simulations agree well with each other, which indicates that our APM can be roughly regarded as a collection of domains of plane wave sources with the designed phases. After being generated by the piezoelectric metasurface, the acoustic waves propagate forward with four equivalent directions of maximum magnitude, as shown in the lower panels of Figure 2c. The directions of maximum amplitude are given by a polar angle of $\theta = \arcsin[\lambda/\sqrt{2}L]$ and azimuthal angles of $\varphi = \pi/4, 3\pi/4, 5\pi/4, 7\pi/4$, where $\lambda$ is the wavelength and L =1.5 mm represents the size of each pixel[21]. Other diffractions can be further suppressed as we increase the number of pixels. The polar angle $\theta$ can be tuned, i.e., beam steering, with the working frequency, as shown schematically in the lower panels of Figure 2c, where from left to right, the working frequencies are 1 MHz, 1.5 MHz and 2 MHz, respectively. Different propagation directions lead to different field distributions at a certain distance, as shown in the upper panels of Figure 2c, where the amplitudes of the pressure fields are measured at $z$ =5 cm above the APM. With increasing frequency, the distance between adjacent hot spots (*d*) decreases. This can be seen more clearly in Figure 2d, wherein we plot both the measured and simulated *d* as a function of frequency. By changing the working frequency, we can realize acoustic beam steering along one direction, and full two-dimensional beam steering can be realized by considering additional tuning parameters or complex coding patterns[21,22]

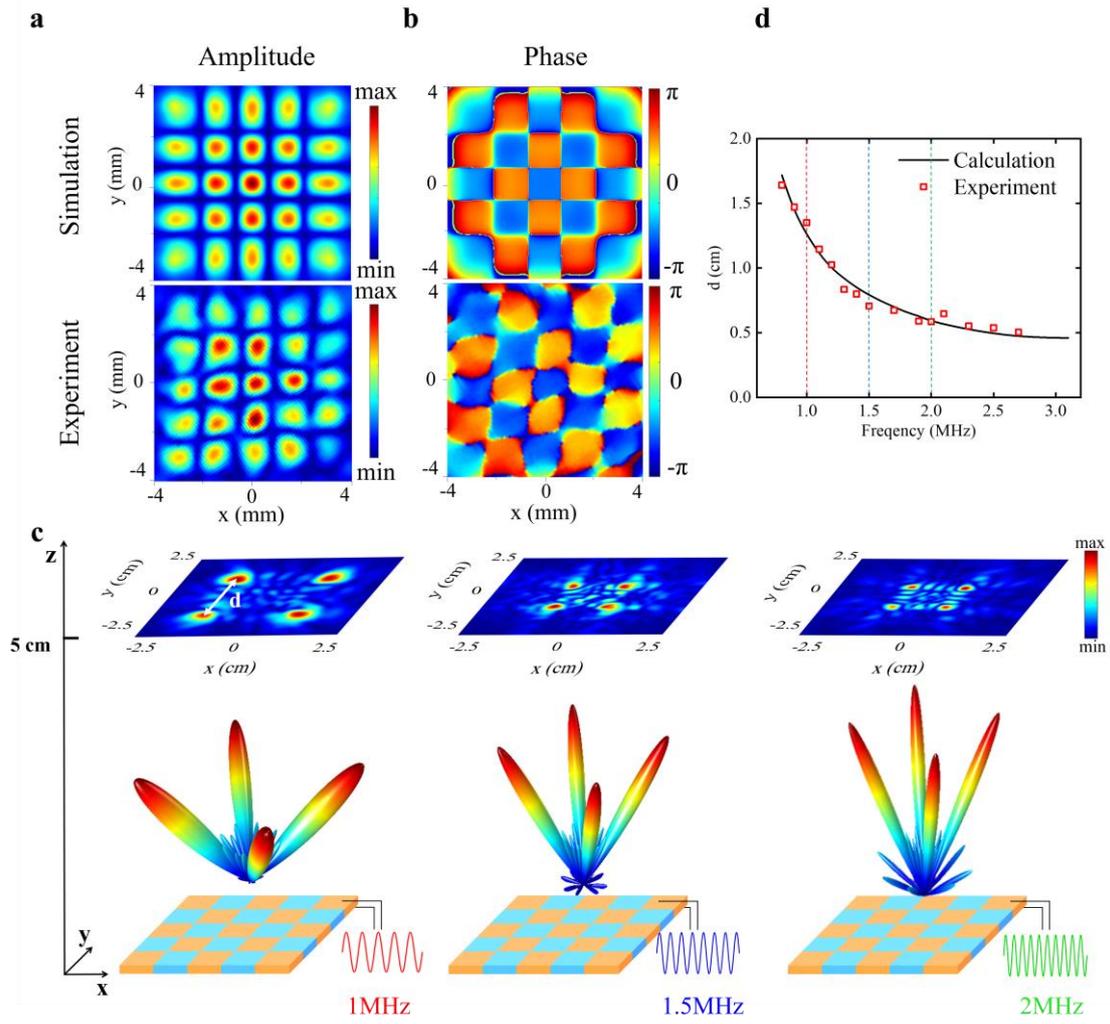

**Figure 2.** Beam steering realized with an APM. **a**, **b**. Simulated (upper panels) and measured (lower panels) pressure field amplitude and phase distributions in the *x-y* plane 2.5 mm above the APM shown in Fig. 1c at a driving frequency of 1.5 MHz. Here, the amplitude distributions are normalized with respect to the corresponding maximum values. **c**, The lower panels show the simulated far-field angular distributions, and the upper panels show the measured pressure field amplitudes at 5 cm above the APM. From left to right, the driving frequencies are 1 MHz, 1.5 MHz and 2 MHz, respectively. **d,** Distance between adjacent hot spots (defined in **c**) as a function of frequency, where the open red squares and the black line represent the measured and simulated results, respectively. The thickness of the APM is 1 mm, the pixel size is 1.5×1.5 mm$^2$, and there are 5 pixels along both the *x* and *y* directions.

## 2.3 Beam Focusing

Sound focusing is one of the most important functionalities for the manipulation of underwater ultrasonic waves[23] and can be realized with acoustic lenses, metasurfaces or PATs[24,25]. However, the sophisticated large-area structures generally required by these techniques inevitably hinder their application in practice. As sketched in **Figure 3**a, here, we show that our APM can generate sharp autofocusing ultrasound with a properly designed polarization distribution. We adopt a modified Airy distribution, which requires the pressure at the source plane to be radially symmetric and described by:

$$p_0(r) = \text{sgn}\left[\text{Ai}\left(\frac{r_0-r}{w}\right)\right]\exp\left(\alpha\frac{r_0-r}{w}\right), \quad (1)$$

where Ai(·) denotes the Airy function, $r$ is the radial distance, $r_0$ is the initial radius of the primary ring, and $w$ is a scale factor with units of length. Here, $\alpha$ is an exponential decay factor to ensure that the wave carries finite energy. Typically, $\alpha$ is small to ensure that the wave behaviour approximates that of an ideal diffraction-free Airy wave packet. Instead of the originally proposed Airy distribution [without the sgn function in Eq. (1)], here, we implement a modified version with our APM; this modification has already been proven to have almost the same efficiency as the original one[26]. Compared with other traditional approaches, such as acoustic lenses, here, the sharp autofocusing effect comes from the unique transverse self-acceleration, which results in high intensity at precisely the focus point with subdiffraction resolution and low intensity elsewhere[10]. The focal point, the width of the profile and the intensity at the focal point are highly tuneable with the control parameters {$r_0$, $w$, $\alpha$}.

To ensure a sharp autofocusing effect, we set $r_0 = 2\lambda$, $w = \lambda$ and $\alpha = 0.05$, where $\lambda = 0.75$mm (frequency: 2 MHz) is the target wavelength of the ultrasonic sound. The resulting phase (red) and amplitude (blue) distributions as functions of $r$ are shown in Figure 3b. The amplitude is uniform, and the phase is either 0 or π. Here, we truncate the function $p_0(r)$ at $r_m = 25$mm, and the contribution from a radial distance beyond $r_m$ is very small since the phase alternates quite rapidly outside. Figure 3c shows a photograph of the APM, where the dark and light golden rings represent the regions where the ultrasound phases are 0 and π, respectively. To test the isotropy of this APM, we measured the $d_{33}$ values for different rings in four representative directions (see the Methods and the Supplementary Materials, Sec. 1). For the regions with phases of 0 and π, the $d_{33}$ values are 247.07 ± 8.04 [pC/N] and 241.15 ± 14.7 [pC/N], respectively, where the first number represents the average value and the latter is the standard deviation.

An APM with a thickness of 1 mm was then assembled as a transducer (see the flow chart presented in the Supplementary Materials, Sec. 2). Figure 3d shows the simulated amplitude of the pressure field on the $y = 0$ plane (50 mm × 30 mm) above the transducer, where $z = 0$ corresponds to the surface of the transducer. The inset in Figure 3d shows the measured amplitude of the pressure field zoomed in around the focal point (the white dashed box), which is quite similar to the simulation. It is clear that the acoustic wave is focused sharply at the focal point of $z = 16.5$ mm ($z = 16.9$ mm in the simulation). Figure 3e presents the simulated (left panels) and measured (right panels) pressure fields on the *x-y* plane at the focal plane. The measured field shows good isotropy in the *x-y* plane, which agrees well with the simulation. Due to background noise, scattering from the boundaries and heterogeneity of polarization, the local minimums of the field (dark black rings in the upper panels) in the experiment are not as sharp as those in the simulation. A detailed comparison between the measured and simulated field amplitudes along the *x*-axis on the focal plane is provided in the Supplementary Materials, Sec. 4, from which we extract the full width at half maximum (FWHM) to be D = 0.63λ in the simulation and D = 0.65λ in the experiment.

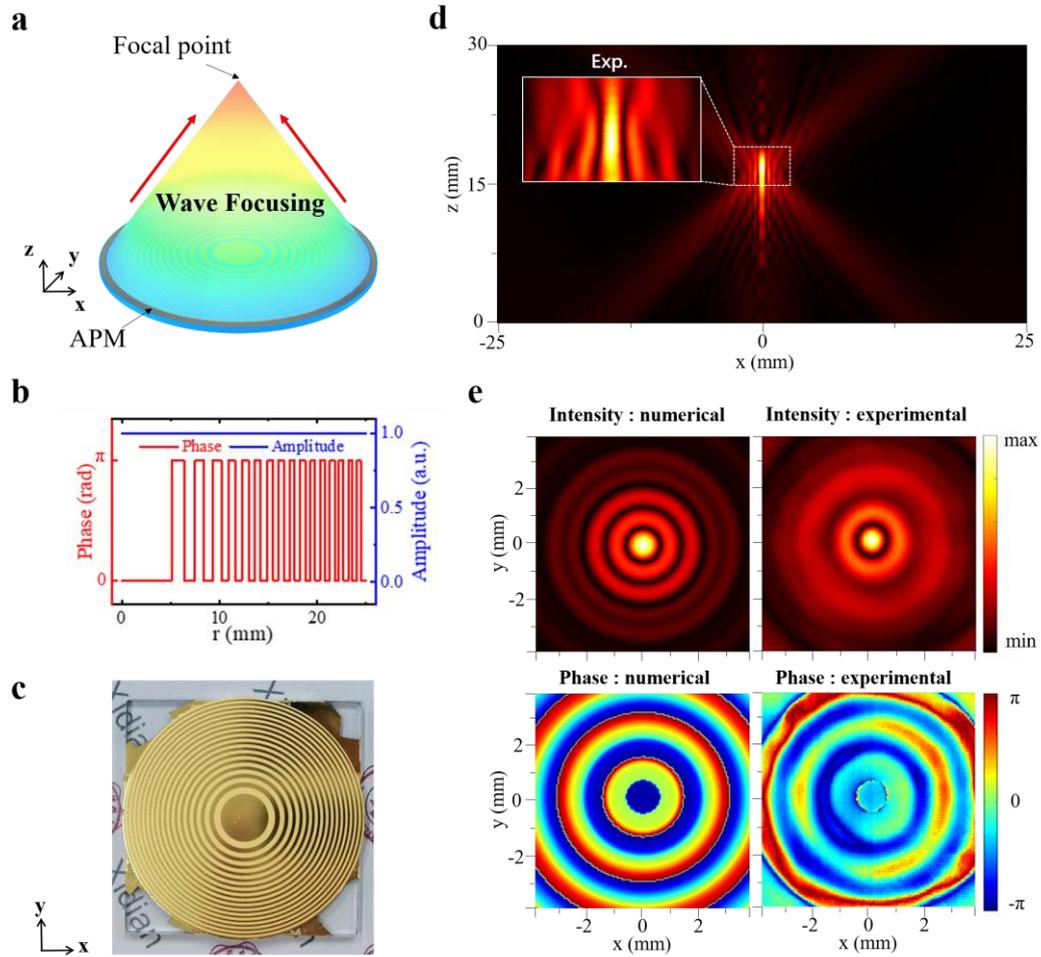

**Figure 3**. Focusing of ultrasonic waves realized with an APM. **a**, A sketch showing the focusing of acoustic waves with a centrally symmetric APM. **b**, The designed amplitude and phase distributions as functions of the radial distance *r*. **c**, A photograph of the APM after photolithography. The dark golden regions consist of prepoled bare piezoelectric material, and the light golden regions are coated with gold, which will be later poled with opposite piezoelectric polarizations. **d**, Simulated amplitude of the pressure field on the *y* = 0 plane above the APM, where *z* =0 corresponds to the surface of the ultrasonic transducer. The inset shows the measured field around the focal point, as marked by the white dashed rectangle. **e**, Simulated (left) and measured (right) pressure amplitude and phase distributions on the *x-y* plane at the corresponding focal planes (*z* =16.9 mm in the simulation and *z* = 16.5 mm in the experiment). In **d** and **e**, the amplitude distributions are normalized with respect to the corresponding maximum values. The radius and thickness of the APM here are 25 mm and 1 mm, respectively.

## 2.4 Vortex Beam Focusing

Acoustic vortex beams have recently attracted considerable attention[27,28]. When an acoustic vortex beam interacts with an object, angular momentum is transferred between the vortex beam and the object. Focusing of a vortex beam can greatly enhance the interaction between the vortex beam and small particles, and thus, the particles can be trapped and rotated simultaneously[29,30,31]. Below, we describe the use of our APM to realize this functionality. Instead of centrosymmetric gratings, we use a spiral diffraction grating, wherein the interferences between the 0- and π-phase branches lead to the formation of a focused vortex beam. Here, we adopt a Fresnel-spiral diffraction grating[32], in which the spiral curves defining the boundaries of the $m$-th arm are given by:

$$r_1(\theta)^2 = \left[\sqrt{r_0^2 + F^2} + \left(\frac{M\theta}{2\pi} - m\right)\lambda\right]^2 - F^2 \qquad (2)$$

$$r_2(\theta)^2 = \left[\sqrt{r_0^2 + F^2} + \left(\frac{M(\theta+\pi)}{2\pi} - m\right)\lambda\right]^2 - F^2 \qquad (3)$$

where $F$ is the focal length in the limit of an infinite grating area, i.e., $r \to \infty$; $\lambda$ is the wavelength; $M$ (set to 1 in this experiment; hence, $m = 1$) is the topological charge of the sound vortex; and $r_0 = \sqrt{(F+\lambda)^2 - F^2}$ is defined as the radius of the first inner spiral circle. With such a Fresnel-spiral grating, waves diffracted from opposite angles arrive at the focal point with opposite phases, which leads to the formation of a focused vortex beam.

**Figure 4**a shows the grating pattern described by Eq. (2) with $\lambda = 1.5\text{mm}$ (frequency: 1 MHz) and $F = 30$ mm. The grating pattern is truncated at $r_m = 25\text{mm}$, with 14 circles; with this truncation, the grating already demonstrates sufficient focusing of the vortex beam. To generate the grating pattern, we first drew two adjacent spiral curves, shown as $r_1(\theta)$ (black dashed line) and $r_2(\theta)$ (phase-demarcated spiral line) in Figure. 4b, and then expanded the spiral curves into arms with phases of 0 (cyan region in Figure. 4b) and π (orange region in Figure. 4b). The amplitude was set to be uniform except inside the region $r < r_0$, wherein the field amplitude is zero (grey region in Figure. 4b). The zero-amplitude region was realized by means of a top electrode possessing a hole with the designed shape; thus, this region is not driven by the input circuit (see the details in the Supplementary Materials, Sec. 2). In contrast to the beam-focusing APM, here, the grating pattern is not centrosymmetric, and the width of the arms decays more slowly. Figure 4c shows the APM after photolithography, where, as before, the dark and light golden colours represent regions where the ultrasound phases are 0 and π, respectively. The APM, with a thickness of 2 mm, was then poled following the same procedure as before, and a characterization of the poling quality is provided in the Supplementary Materials, Sec. 1. As before, this APM was assembled as a standard ultrasonic transducer after poling.

Figure 4d shows the simulated amplitude of the pressure field on the $y = 0$ plane above the transducer, with the top surface of the transducer at $z = 0$ mm. The inset in Figure 4d shows the measured amplitude of the pressure field zoomed in at the focal point (white dashed box), which is similar to the simulation. Similar to Figure 3d, the pressure field is focused at $z = 20.1$ mm, which is different from $F = 30$ mm because of the finite area of the Fresnel-spiral grating.[33] To better characterize the vortex property of the focused beam, we show the simulated (left) and measured (right) pressure field distributions for both the amplitude (upper panels) and phase (lower panels) on the focal plane in Figure 4e. From Figure 4e, we can clearly see a minimum amplitude at the centre and a phase vortex around it. The vortex beam realized here is not isotropic since the structure is already not centrosymmetric. Note that here, we consider a vortex beam with topological charge $M = 1$; however, other vortex beams with higher topological charges could also be generated following the same procedures.

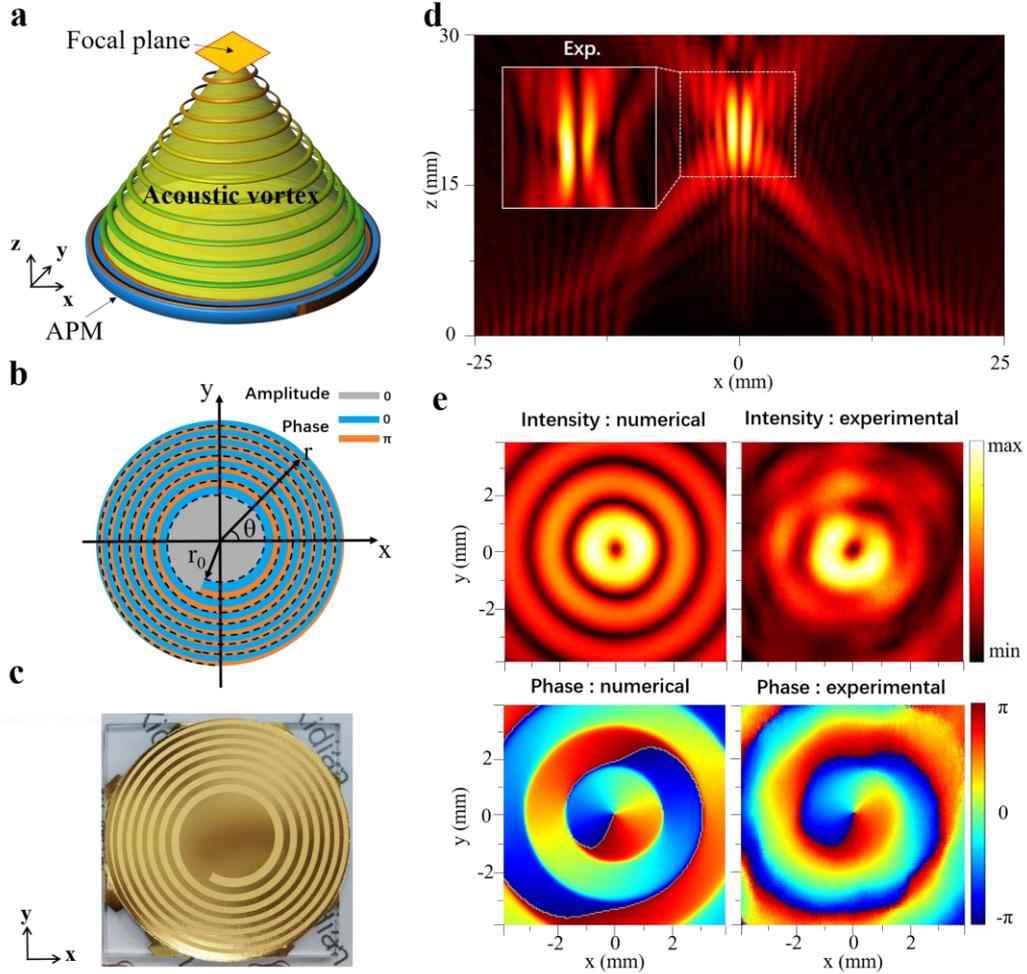

**Figure 4**. Focusing of a vortex beam with an APM. **a**, A sketch showing the focusing of an acoustic vortex beam. **b**, The grating pattern of the designed APM, where grey represents the region with zero amplitude and cyan and orange represent regions with uniform amplitude and phases of 0 and π, respectively. **c**, A photograph of the APM after photolithography, where dark and light golden colours correspond to regions of bare PZT-4 and PZT-4 coated with gold for later poling, respectively. **d**, Simulated and measured (inset) amplitudes of the pressure field on the $y = 0$ plane, where the measured region corresponds to the white dashed rectangle. **e**, Simulated (left) and measured (right) amplitude (upper) and phase (lower) distributions of the pressure fields on the corresponding focal planes. Here, the amplitude distributions are normalized with respect to their corresponding maximum values, the working frequency is 1 MHz, and the radius and thickness of the APM are 25 mm and 2 mm, respectively.

## 2.5 Particle Manipulation with a Focused Beam Generated by an APM

Contact-free trapping and precise manipulation of small particles and cells are of vital importance in various kinds of biological applications, such as bioengineering, cell research, drug screening and neural regulation[34,35,36]. Compared with optical tweezers, ultrasonic tweezers are more suitable for manipulating objects with a size of approximately a few hundred µm while also being biologically safe[37,38]. Here, we show that the focused beam generated by the APM in Figure 3 can be used as ultrasonic tweezers for trapping and manipulating microbubbles. The ability of our APM to focus vortex beams is investigated in the Supplementary Materials, Sec. 6. For illustration, we first numerically calculate the force distribution on the focal plane (z=19.9 mm) of the APM (see the Methods), as shown in **Figure 5**a. The white arrows show the convergence of the force field towards the centre of the beam, indicating that the focused beam can indeed trap microparticles. Since the focused beam is isotropic, the trapping potential is also isotropic. Figure 5b shows the normalized lateral force along the $x$-axis (marked as a white dashed line in Figure 5a). Around the focal point (red point), the lateral force is negative for $x > 0$ and positive for $x < 0$, and the radius of the trapping region (defined as the FWHM of the focused beam) is 550 µm.

Figure 5c and 5d show a structure diagram and a photograph, respectively, of the setup for our ultrasonic tweezers. The ultrasonic transducer made from the APM in Figure 3 is hung from an adjustable bracket and immersed in deionized water in a transparent glass chamber. The glass chamber is fixed on the plate of a motorized stage, which can then be moved to demonstrate the capabilities of the acoustic tweezers. Small particles made of polystyrene are placed on the bottom of the glass chamber. An inverted microscope is used to record the trajectory of these particles. Figure 5e shows selective manipulation of a particle with a radius of approximately 150 µm (highlighted with yellow dashed circles) using our tweezers. Here, movements along three representative directions, as represented by pink and white dashed arrows, are shown (see also Supplementary Movie I), demonstrating the ability to trap and move particles along an arbitrary direction. Moreover, our acoustic tweezers with a focused beam are able to pick up a single particle with a radius of approximately 150 µm without affecting other surrounding particles (see also Supplementary Movie II). The focused vortex beam generated by our APM can also be used as acoustic tweezers, with a trapping region (defined as the FWHM of the vortex beam) of approximately 500 µm (see the

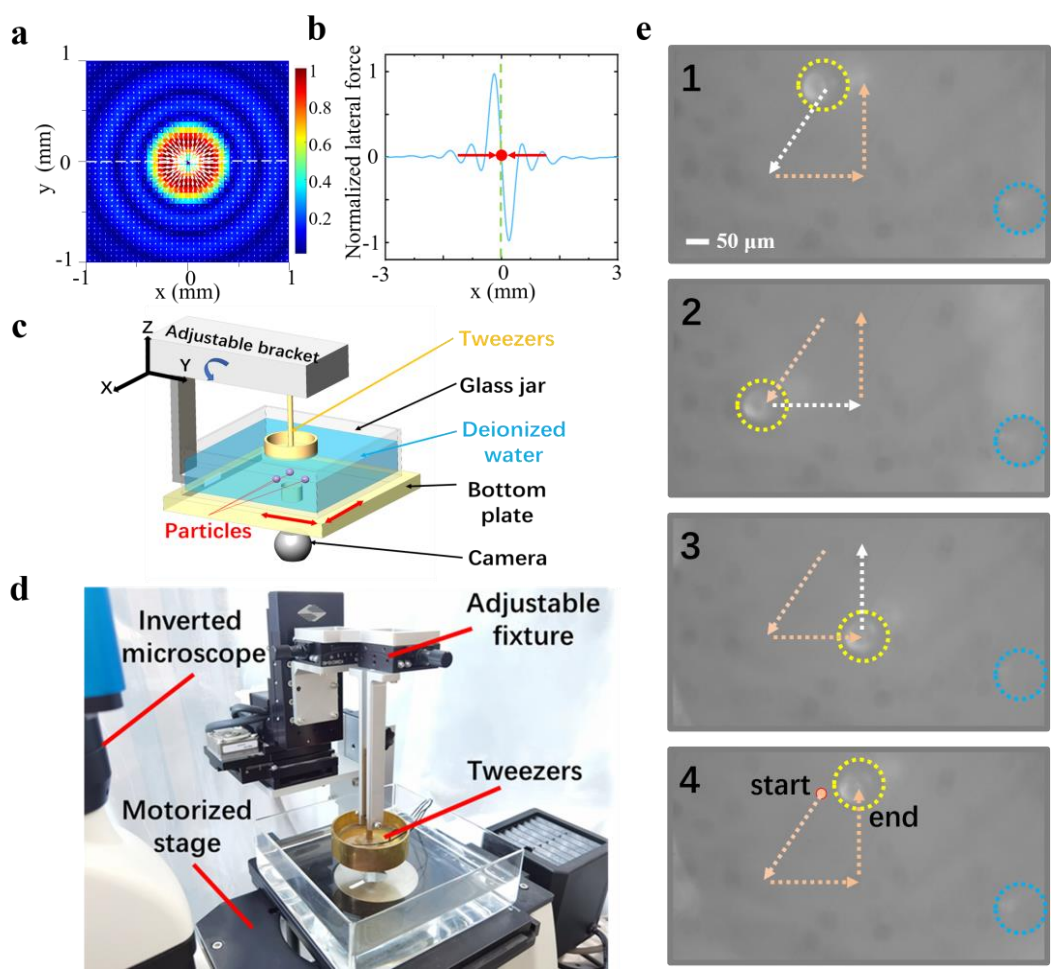

**Figure 5**. Acoustic tweezers realized with an APM. **a**, Acoustic force distribution at the focal plane of the focused beam in Figure. 3, where the white arrows indicate the directions and the lengths of the arrows and the colour of the plot represent the magnitude of the force. **b**, Magnitude of the lateral force along the white dashed line in **a**. **c,** Structure diagram of the acoustic tweezers. **d**, A photograph of the experimental setup. **e**, Snapshots show the trapping and movement of a particle (marked with a yellow dashed circle) with a radius of approximately 50 µm, where the pink dotted arrows show the trajectories of this particle. The cyan dashed circles indicate another reference particle, which is unaffected by the manipulation.

## 2.6 Ultrasonic imaging application of the APM

Ultrasound imaging serves as an important role in both clinical use and non-destructive testing due to its non-toxicity, low-cost and practical flexibility. The imaging technologies including tissue characterization and image segmentation, micro scanning and intravascular scanning, elastic imaging, reflection transmission imaging, computed tomography, Doppler tomography, photoacoustics and thermoacoustics are developing rapidly.

Here we show the imaging applications of the proposed APM in both water ambient and tissue-approximating phantoms, Tissue-approximating phantoms can be used as a substitute for human tissue for its similar acoustic characteristics of human tissue. The tissue-like phantoms were made following the procedures in Supplementary Materials, Sec. 9. **Figure 6**a. shows the structure diagram of the experimental setup in water ambient (see the Methods). Figure 6c. are the results of the B-Mode imaging (lateral scan) of 1.5 mm diameter tungsten rods with gaps of 0.5 mm in water ambient with an APM. Figure 6e. is the intensity map of the APM after the lateral scan of 1.5 mm diameter tungsten rods with gaps of 0.5 mm in water ambient, the intensity is normalized to the maximum value. Obviously, the lateral resolution for the APM is capable to distinguish two tungsten rods clearly. The APM can clearly distinguish two tungsten rods with gaps of 0.5 mm in water ambient as the signal shows a sharp decrease at the edges of each rod, the two white areas near the center in Figure 6c. are artifacts due to the experimental error. The lateral resolution of ultrasound imaging is mainly determined by the width of the focused beam. Using the half of two peak's intensity(FWHM) to determine the resolving power and the FWHM of the APM is 0.52 mm (left rod) and 0.49 mm (right rod). Figure 6b. shows the structure diagram of the experimental setup in tissue-approximating phantoms (see the Methods). Figure 6d. is the results of the B-Mode scan (lateral scan) of 1.5 mm diameter copper rods in tissue-approximating phantoms. Figure 6f. is the intensity map of the APM after the lateral scan of 1.5 mm diameter copper rods in tissue-approximating phantoms. As we can see in Figure 6f. the FWHM of the APM 0.51mm. In order to further demonstrate the ultrasound imaging capability of APM, C mode ultrasound imaging experiment were carried out. Using a customized Scanning Acoustics Microscope system (see the Methods), the APM was utilized to image the RMB one yuan coin. Figure 6g. shows the structure diagram of the experimental setup(same as a) of the C mode ultrasound imaging in water (see the Methods). A coin was stick to the adjustable fixture platform underwater, the focal plane of the APM was set on the surface of the coin. Figure 6h. was the gray scale image of a coin after C mode ultrasound imaging of the APM. The dynamic range of the gray scale image is 40 dB. The C-scan image of the coin shows high resolution and good signal-to-noise ratio, which proves the excellent imaging capability of the APM. In short, the APM has a high resolution of ultrasound imaging due to its good focusing effect both in the water and tissue-approximating phantoms, which provides a new method for the application of ultrasound medical imaging.

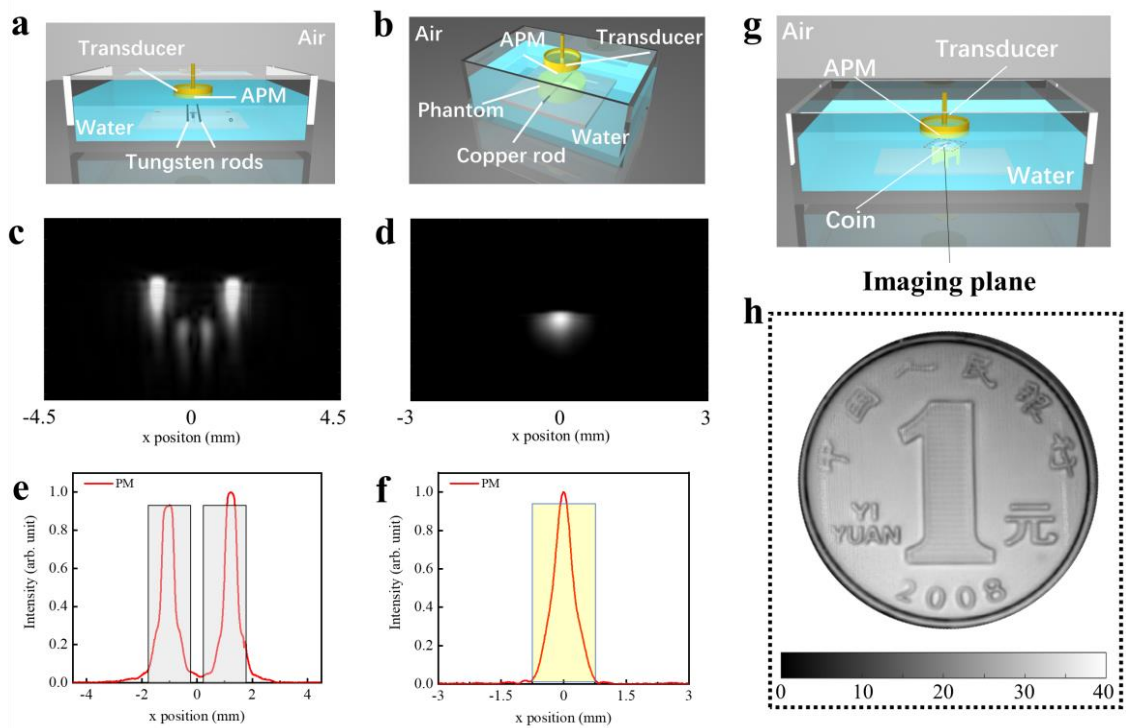

**Figure 6**. The performance of the APM in both water ambient and tissue-approximating phantoms. **a,** structure diagram of the experimental setup in water ambient (details in Supplementary Materials, Sec. 7). **b**, structure diagram of the B mode ultrasound imaging experimental setup in tissue-approximating phantoms (details in Supplementary Materials, Sec. 7). B-mode imaging results of the APM (**c**) in water ambient. **e**, the intensity map of the APM after the scan of 1.5 mm diameter tungsten rods with gaps of 0.5 mm in water ambient, the intensity is normalized to the maximum value. B-mode imaging results of the APM (**d**) in tissue-approximating phantoms. **f**, the intensity map of the APM after the scan of 1.5 mm diameter copper rods in tissue-approximating phantoms. **g**,

structure diagram of the experimental setup(same as **a**) of the C mode ultrasound imaging in water. **h**, the gray scale image of a coin after C mode ultrasound imaging of the APM.

## 2.7 Sub-wavelength imaging of the APM

Experiments are carried out to further demonstrate the sub-wavelength ultrasound imaging ability of the APM in water. The experimental setup is given in **Figure 7**a (see Methods and Supplementary Materials, Sec. 8). The stainless steel plate with different patterns is placed in the focal plane (x-y) of the APM, which is moved in the x-y plane by a three-dimensional precision moving stage. A hydrophone close to the lower bottom of the stainless steel plate is used to scan the pressure field.

Figure 7b shows the 0.2 mm thick stainless steel plate with different patterns. (Structure parameters are given in Supplementary Materials, Sec. 8 ). A hole array and a helix line are designed to check the imaging ability of APM. The diameter of the biggest and smallest holes are 1.2 mm and 0.5 mm, the width of the helical line is 0.5 mm. Figure 7c is the image result of the APM, which shows all major features of the patterns clearly. The gap size and width of 5 groups of double slits are 100 μm (0.13$\lambda$), 200 μm (0.27$\lambda$), 300 μm (0.4$\lambda$), 400 μm (0.53$\lambda$), and 500 μm (0.67$\lambda$), with $\lambda$=750 μm at 2 MHz, Figure 7d and e are the sub-wavelength images of the double slits with different gaps. The intensity maps across the center of 5 groups of double slits are shown in Figure 7f and g, demonstrate the double slits with a sub-wavelength lateral gap of 300 μm (0.4$\lambda$) can be distinguished, which shows good sub-wavelength imaging results.

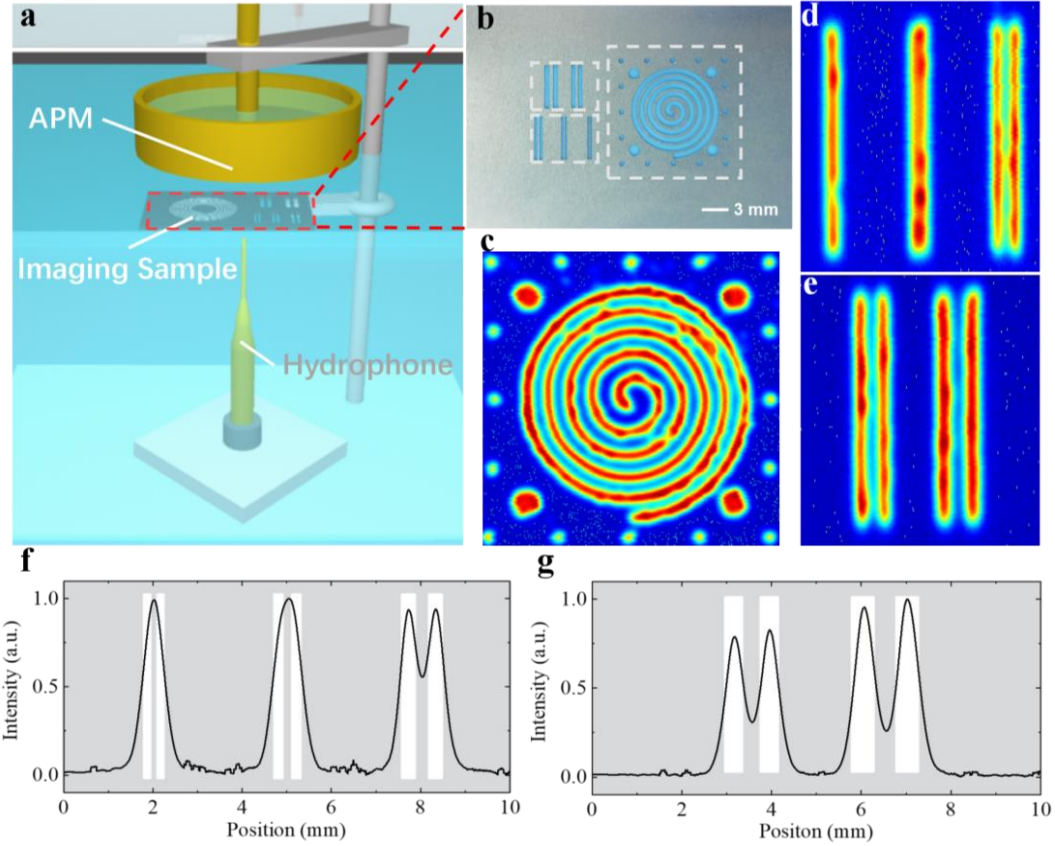

**Figure 7.** Sub-wavelength acoustic imaging. **a,** the sub-wavelength imaging experimental setup. **b,** the diagram of the stainless steel plate with patterns. **c,** the ultrasound images of a helical line and hole array. **d** and **e** are the sub-wavelength images of double slits with different gaps. The pressure fields in **c**, **d** and **e** are normalized to the maximum of each of those plots. **f** and **g** are the intensity maps across the center of double slits in **d** and **e**. Intensity is expressed in arbitrary units (a.u.).

## 3. Discussion

In this work, we have experimentally demonstrated the binary phase information coding of several APMs based on the polarization of piezoelectric materials. Our APMs are able to perform beam steering and generate focused beams and focused vortex beams with a single electrode input, and the focused beams and focused vortex beams thus generated can be further used as acoustic tweezers for capturing and manipulating microparticles. In contrast to previous approaches based on metasurfaces or PATs, our scheme relies on frequency-independent phase information coding and requires only a single input electrode and no other external accessories. Such a simple and robust scheme is quite suitable for large-scale production and presents broad possibilities for acoustic-wave modulation and ultrasound applications. Note that although we have only demonstrated the functionalities of our APMs for ultrasonic underwater sound waves, our approach also works over a much broader frequency range and under different circumstances, such as for airborne sound and elastic waves. Moreover, the resolution of our APMs is similar to that of lithography, so the working frequency of this type of APM can, in principle, reach tens or even hundreds of megahertz.

## 4. Experimental Section

*Simulations of Acoustic field generation by APMs.* The acoustic fields generated by the APMs were simulated with COMSOL Multiphysics[28]. The generation of the acoustic waves in Figure 2a and 2b was modelled using the Frequency Domain setting of the Pressure Acoustic module (the structure is shown in the Supplementary Materials, Sec. 3). The water was modelled as an acoustic domain with radiation boundary conditions applied to the external boundaries. The APM therein was modelled as an array of plane wave sources with a uniform amplitude and a phase of 0 or $\pi$ to approximately represent the domains with opposite directions of polarization. The far-field angular distributions in Figure 2c were obtained using the method described in Ref. [20]. The generation of the acoustic waves in Figure 3 and 4 was simulated using the Frequency Domain setting of the Coupled Elastic-Acoustic module (the structure is shown in the Supplementary Materials, Sec. 3). Similarly, the water was modelled as an acoustic domain with radiation boundary conditions applied to the external boundaries. The contact surface between the water and the piezoelectric material PZT-4 was modelled as a multiphysics coupling of the acoustic structure boundary. The polarization direction of the PZT-4 was modelled using the opposite base vector coordinate system. The generation of sound waves by the PZT-4 was modelled using the Solid Mechanics and Electrostatics modules in combination with a piezoelectric multiphysics coupling. A sinusoidal electric potential difference with the selected frequency was applied at the upper and lower surfaces of the APM in the Electrostatics module. To absorb the backward propagating waves, suppress other vibration modes and provide mechanical support for the APM, we also added a 5 mm thick epoxy backing layer below the APM. The epoxy backing layer was modelled using the Solid Mechanics module of COMSOL. The parameters used for the PZT-4, epoxy backing layer and water are provided in the Supplementary Materials, Sec. 3.

*Calculation of radiation force.* The radiation force acting on an elastic sphere was calculated using the method developed by Sapozhnikova[39]. That work offers a theoretical approach for calculating the radiation force on an elastic sphere subjected to an arbitrary acoustic beam. The three components of the radiation force ($F_x$, $F_y$, and $F_z$) are given by:

$$F_x = \frac{1}{8\pi^2 \rho c^2 k^2} \text{Re}\left\{ \sum_{n=0}^{\infty} \psi_n \sum_{m=-n}^{n} A_{nm}(H_{nm}H^*_{n+1,m+1} - H_{n,-m}H^*_{n+1,-m-1}) \right\}, \quad (4)$$

$$F_y = \frac{1}{8\pi^2 \rho c^2 k^2} \text{Im}\left\{ \sum_{n=0}^{\infty} \psi_n \sum_{m=-n}^{n} A_{nm}(H_{nm}H^*_{n+1,m+1} + H_{n,-m}H^*_{n+1,-m-1}) \right\}, \quad (5)$$

$$F_z = \frac{1}{8\pi^2 \rho c^2 k^2} \text{Re}\left\{ \sum_{n=0}^{\infty} \psi_n \sum_{m=-n}^{n} B_{nm}H_{nm}H^*_{n+1,m} \right\}. \quad (6)$$

Here, $H_{nm} = \iint_{k_x^2+k_y^2 \leq k^2} dk_x dk_y S(k_x,k_y) Y^*_{nm}(\theta_k,\varphi_k)$, where $S(k_x,k_y)$ is the angular spectrum of the acoustic waves generated by the transducer; $k$ is the wavenumber in water; $\theta_k$ and $\varphi_k$ are the azimuthal angle and polar angle, respectively, of the wave vector $\mathbf{k} = (k_x, k_y, \sqrt{k^2-k_x^2-k_y^2})$; $c$ is the sound speed in water; $\rho$ is the density of water; and the $Y_{n,m}(\theta_k,\phi_k)$ are spherical harmonics. $A_{nm} = \sqrt{(n+m+1)(n+m+2)/(2n+1)(2n+3)}$, $B_{nm} = \sqrt{(n+m+1)(n-m+1)/(2n+1)(2n+3)}$, and $\psi_n = 2(c_n + c^*_{n+1} + 2c_n c^*_n)$, with the $c_n$ being the coefficients of the spherical waves scattered from a nonabsorbent isotropic elastic sphere, as defined in Ref. [39].

*Surface pattern lithography of APMs.* A device with the required Cr/Au patterns was fabricated with metal lift-off technology. First, a URE-2000S/25 system was used to perform the photolithography process, and a layer of patterned photoresist (ALLRESIST, AR-P 3510T) with a thickness of 2.5 μm was used as a mask. Then, a 50 nm Cr film and a 100 nm Au film were successively coated using an RF-magnetron sputtering (JGP560 system, Sky Technology Development Co., Ltd., Chinese Academy of Sciences). Finally, the photoresist mask was lifted off in acetone, leaving the required Cr/Au pattern on the piezoelectric substrate.

*Poling of APMs.* The poling setup is shown in Supplementary Figure S1. The piezoelectric materials were prepoled before photolithography to ensure that the polarization direction would be upward. The region that needed to be poled with downward polarization was covered by the Cr/Au pattern deposited via lithography. The patterned piezoelectric materials were immersed in high-temperature silicone oil, which was kept at 60 °C by a thermostat heater (DF-101S, YUHUA, China). A copper sheet (cathode) was connected to the bottom surface of the piezoelectric material as the ground plane, and a copper needle (anode) was used to select the regions that needed to be poled on the upper surface of the material. A high voltage difference applied between the cathode and anode was provided by a DC power supply system (PS350/5000V-25W, Stanford Research Systems, USA). Each ring was poled for 5 minutes, and the applied voltage was gradually increased from 0 to 2500 V, with a maximum value above the coercive field of PZT-4.

*Measurement of $d_{33}$ values of APMs*. To test the quality of the poling process, we measured the $d_{33}$ values (ZJ-4AN, The Institute of Acoustics of the Chinese Academy of Sciences, China) of the APMs in different domains. Details can be found in the Supplementary Materials, Sec. 1. For the APM in Figure 2, we measured the $d_{33}$ value of each of the 25 domains of the APM. The $d_{33}$ values for the regions with phases of 0 and π were 236.4 ± 8.25 [pC/N] and 235.0 ± 7.22 [pC/N], respectively, where the first number represents the average value and the second is the standard deviation. For the APMs in Figures 3 and 4, we chose four representative directions and measured $d_{33}$ for each ring. For the APM in Figure 3, the $d_{33}$ values for the regions with phases of 0 and π were 241.2 ± 14.71 [pC/N] and 221.8 ± 11.60 [pC/N], respectively. For the APM in Figure. 4, the $d_{33}$ values for the regions with phases of 0 and π were 247.1 ± 8.72 [pC/N] and 250.4 ± 9.98 [pC/N], respectively.

*Assembly of ultrasound transducers*. A flowchart of the assembly process is provided in the Supplementary Materials, Sec. 2. As shown in Figures. S6b and S7b, on the upper surface of the APM (the surface on which the Cr/Au pattern was deposited), another layer of Cr/Au (50/100 nm) was first deposited and connected out by copper wires with E-solder (VonRoll Isola, New Haven, CT). For the APMs in Figure 2 and 3, this new Cr/Au layer covered the whole surface. For the APM in Figure 4, this new Cr/Au layer covered the whole surface except for the grey region in Figure. 4b, with the help of another properly designed mask. The wired APM was then placed in a Cu housing (Figures. S6c and S7c), and a layer (thickness: 5 mm) of epoxy resin (American Safety Technologies, Roseland, NJ) was poured onto it to provide mechanical support and reduce clutter vibration (Figures. S6d and S7d). Subsequently, the copper housing was heated to 60°C for 2 hours to cure the epoxy. Then, another Cr/Au (50/100 nm) layer was sputtered onto the lower surface of the APM and the Cu housing (Figures. S6e and S7e) such that the Cu housing was electrically connected to the lower surface of the APM and could act as another electrode. The whole system was then coated with parylene as a protective layer (Figures S6f and S7f).

*Measurement of pressure fields.* The pressure fields generated by the transducers were measured with a homemade scanning system based on LabVIEW (the details are shown in the Supplementary Materials, Sec. 6). The transducer to be measured was connected to a JSR Ultrasonics DPR 500 (Imaginant, NY) pulser/receiver and excited by an electrical impulse at a 200 Hz repetition rate and 50 Ω damping. The energy was 2.3 μJ per pulse, and no gain was applied. A hydrophone (NH1000, PA, UK) with a size of 50 μm was used to scan the pressure field. The signal received by the hydrophone was recorded in the time domain by a data acquisition card (QT1140, Queentest, China) with a sampling rate of 125 MHz. The recorded time-domain data were then Fourier transformed in the processing system, and the amplitude and phase of the pressure field were obtained as functions of frequency.

*Manipulation of microparticles with acoustic tweezers.* The structure and other details of the tweezer system are presented in the Supplementary Materials, Sec. 5. The tweezers (those made with both the beam-focusing APM and the vortex-beam-focusing APM) were set in a rectangular chamber filled with distilled water and mounted on a motorized three-axis positioner (ROE-200, Sutter Instrument, USA). A complementary metal–oxide–semiconductor (CMOS) camera (XCAM1080P, Jingtong, China), in combination with an inverted microscope (DM43T, LIOO, Germany), was placed under the chamber to record the motion of the trapped sample. Initially, the position of the acoustic tweezers was adjusted to focus at the bottom of the chamber using the pulse-echo method[40]. A function generator (33120A, Hewlett-Packard) together with a 50 dB power amplifier (535LA, ENI, Rochester, MN) were used to drive the transducer. The operating frequencies of the focused-beam and vortex-beam tweezers were 2 MHz and 1 MHz, respectively. The microparticles were polystyrene microspheres (Monodisperse Polystyrene Microspheres, PD1281, Rigor, China) with a diameter of approximately 100 μm. The particles had a density of 1050 kg m$^{-3}$, and the speeds of the longitudinal and transverse waves were 2340 m s$^{-1}$ and 1100 m s$^{-1}$, respectively.

*Measurements of the ultrasound B mode imaging of the APM.* The APM/transducer was connected to a JSR Ultrasonics DPR 500 (Imaginant, Pittsford, NY) pulser/receiver and excited by an electrical impulse at 200Hz repetition rate and 50Ω damping. The energy involved was 2.3 μJ and no gain was applied in excitation or reception. APM/Transducers were used to realize the lateral scan of the rods by using a customized ultrasonic bio microscopy (UBM) system. Each 2-D frame of image data was obtained by scanning the APM/transducers using motor controlled by computer and collecting pulse-echo lines using a 200 MHz digital storage oscilloscope (DSOX3024A, KEYSIGHT, USA) at 5 μm spacing. A custom LabVIEW program (National Instruments, Austin, TX, USA) was used to control the imaging procedure and save the raw RF data for offline processing. The imaging depth was determined by the ultrasound penetration depth. A logarithmic compression algorithm was used to improve grayscale visualization of the image.

*Measurements of the ultrasound C mode imaging of the APM.* During the imaging process of scanning ultrasonic microscope, the scanning motion command was sent by the computer, and then, the motion control chassis received and processed the command and sent the motor-driven signal to the scanning motion platform, the scanning motion platform with the APM start to move, and outputs the positioning synchronization pulse to the pulse receiving and transmitting equipment. The pulse transmitting and receiving equipment sent out the excitation signal to excite the APM and transmit the received echo signal to the data acquisition equipment after triggered by the pulse. Finally, the computer carries out data processing and imaging display. The imaging algorithm uses the amplitude integral method (BAI), processing the time-domain echo obtained at each acquisition position to obtain the integral value of each acquisition position by Hilbert envelop methods, and generates a normalization matrix for imaging.

*Measurements of the sub-wavelength ultrasound imaging of the APM.* The Experimental setup is given in the Supplementary Materials, Sec. 8. A 10-cycle sine wave at 2MHz generated by a multi-function signal generator (SMB100A, Rohde & Schwarz, GER) is amplified by a 26 dB power wideband amplifier (A TA-122D, Aigtek, CHN) and then applied to the transducer. The stainless steel plate with different patterns is placed in the focal plane (x-y) of the APM, which is moved in the x-y plane by a three-dimensional precision moving stage (H2-2206, ESM, CN). A hydrophone (NH1000, PA, UK) close to the lower bottom of the stainless steel plate with a size of 20 μm was used to scan the pressure field. The signal received by the hydrophone was recorded in the time domain by a data acquisition card (QT1140, Queentest, China) with a sampling rate of 125 MHz. The recorded time-domain data were then Fourier transformed in the processing system, and the amplitude of the pressure field was obtained.

## Supporting Information

Supporting Information is available from the Wiley Online Library or from the author.

## Acknowledgements

M.X., C.F., and Y.Y. initiated and supervised the project. Z.L. conducted the simulations and designed the samples. Z.L., J.Z., C.H., Y.L., H.W., Y.Q., N.G., and C.Z. performed the experiments. M.X., C.F., and Z.L. analysed the data. M.X., C.F., Z.L. and S.Y. wrote the manuscript. All authors contributed to discussions of the manuscript. M. X. thanks the insightful discussions with Zhengyou Liu. This work was supported by the National Natural Science Foundations of China (Grant No. 51802242, No. 11904264), Shenzhen Science technology and fundamental research and discipline layout project (No. JCYJ20170818153048647), the National Natural Science Foundation of Shaanxi Province (Grant No. 2019JQ-313, 2020JM-205, 2020JM-253), the China Postdoctoral Science Foundation (Grant No. 2019M663927XB), the 111 Project (No. B12026), and the Fundamental Research Funds for the University of People's Armed Police (No: WJY201923, WJY202147).

## Conflict of Interest

The authors declare no conflict of interest.

## Keywords

coding piezoelectric metasurface, opposite polarizations, beam steering, particle manipulation, ultrasonic imaging